\documentclass[twocolumn,showpacs,preprintnumbers,amsmath,amssymb]{revtex4}
\usepackage{graphicx}
\usepackage{epsfig}
\usepackage{dcolumn}
\usepackage{bm}
\newcommand{\psipto}{\psi(2S)\rightarrow \pi^+\pi^- J/\psi}

\newcommand{\EE}{e^+e^-}

\newcommand{\GG}{\gamma\gamma}
\newcommand{\pp}{\pi^+\pi^-}
\newcommand{\ppp}{\pi^+\pi^-\pi^0}

\newcommand{\etap}{\eta^{\prime}}
\newcommand{\gpi}{\gamma\pi^0}
\newcommand{\geta}{\gamma\eta}
\newcommand{\getap}{\gamma\etap}

\newcommand{\psp}{\psi(2S)}
\newcommand{\jpsi}{J/\psi}
\newcommand{\ar}{\rightarrow}

\newcommand{\bfg}{\begin{figure}}
\newcommand{\efg}{\end{figure}}
\newcommand{\bitm}{\begin{itemize}}
\newcommand{\eitm}{\end{itemize}}
\newcommand{\bnum}{\begin{enumerate}}
\newcommand{\enum}{\end{enumerate}}
\newcommand{\btbl}{\begin{table}}
\newcommand{\etbl}{\end{table}}
\newcommand{\btbu}{\begin{tabular}}
\newcommand{\etbu}{\end{tabular}}
\newcommand{\bcl}{\begin{center}}
\newcommand{\ecl}{\end{center}}
\newcommand{\bbt}{\bibitem}
\newcommand{\beq}{\begin{equation}}
\newcommand{\eeq}{\end{equation}}
\newcommand{\beqr}{\begin{eqnarray}}
\newcommand{\eeqr}{\end{eqnarray}}

\begin{document}
\title{\boldmath Search for the C-parity violating process $\jpsi\ar\GG$ via $\psipto$}
\author{
M.~Ablikim$^{1}$,              J.~Z.~Bai$^{1}$, Y.~Ban$^{12}$,
X.~Cai$^{1}$,                  H.~F.~Chen$^{17}$,
H.~S.~Chen$^{1}$,              H.~X.~Chen$^{1}$, J.~C.~Chen$^{1}$,
Jin~Chen$^{1}$,                Y.~B.~Chen$^{1}$
% ...
\\
\vspace{0.2cm}
(BES Collaboration)\\
\vspace{0.2cm}
{\it
$^{1}$ Institute of High Energy Physics, Beijing 100049, People's
Republic of China\\
% ...
$^{12}$ Peking University, Beijing 100871, People's Republic of China\\
% ...
$^{17}$ University of Science and Technology of China, Hefei 230026,
People's Republic of China\\
% ...
}
}
\date{\today}

\begin{abstract}
Using $14.0\times 10^6~\psi(2S)$ events collected with the BES-II
detector, the C-parity violating process $\jpsi\ar\GG$ via $\psipto$
is studied.  We determine a new upper limit for the $\jpsi\ar\GG$
branching ratio of ${\cal B}(\jpsi\ar\GG)<2.2\times 10^{-5}$ at the
90\% C.L., which is about 20 times lower than the previous
measurement.
\end{abstract}

\maketitle
\section{Introduction}
C-parity violation has been studied in different electromagnetic
decays~\cite{cv1,cv2}. 
In this paper, we present a search for the C-parity violating decay,
$\jpsi\ar\GG$. In a previous measurement~\cite{cntr}, the direct
$\jpsi\ar\GG$ decay was used, and the upper limit measured is ${\cal
B}(\jpsi\ar\GG)<5\times 10^{-4}$. The obvious disadvantage of that
measurement is the large QED background from $\EE\ar\GG$. In this
analysis we study this decay via $\psipto,\jpsi\ar\GG$. Therefore, the
QED background can be strongly suppressed since we observe a $\pp$
pair plus two photons and do not base our search just on the two
$\gamma$ invariant mass distribution. As a result, the precision is
significantly improved.

\section{The BES detector and Monte Carlo}
The Beijing Spectrometer (BES) is a conventional solenoidal magnet
detector that is described in detail in Ref.~\cite{bes}; BESII is the
upgraded version of the BES detector~\cite{bes2}. A 12-layer vertex
chamber (VC) surrounding the beam pipe provides trigger and position
information. A forty-layer main drift chamber (MDC), located radially
outside the VC, provides trajectory and energy loss ($dE/dx$)
information for charged tracks over $85\%$ of the total solid angle.
The momentum resolution is $\sigma _p/p = 0.017 \sqrt{1+p^2}$ ($p$ in
$\hbox{\rm GeV/c}$), and the $dE/dx$ resolution for hadron tracks is
$\sim 8\%$.  An array of 48 scintillation counters surrounding the MDC
measures the time-of-flight (TOF) of charged tracks with a resolution
of $\sim 200$ ps for hadrons.  Outside of the TOF counters is a
12-radiation-length barrel shower counter (BSC) composed of gas tubes
interleaved with lead sheets. This measures the energies of electrons
and photons over $\sim 80\%$ of the total solid angle with an energy
resolution of $\sigma_E/E=22\%/\sqrt{E}$ ($E$ in GeV).  Outside of the
solenoidal coil, which provides a 0.4~Tesla magnetic field over the
tracking volume, is an iron flux return that is instrumented with
three double layers of counters that identify muons of momentum
greater than 0.5 GeV/c.

A GEANT3 based Monte Carlo (MC) program with detailed consideration of
the detector performance (such as dead electronic channels) is used to
simulate the BESII detector.  The consistency between data and Monte
Carlo has been carefully checked in many high purity physics channels,
and the agreement is quite reasonable~\cite{simbes}.  For
$\psipto,\jpsi\ar\GG$, the mass of the dipion system is generated 
according to~\cite{fred}
$$\frac{d\sigma}{dM_{\pp}}\propto (phase~space)\times(M_{\pp}^2-4m_{\pi}^2)^2,$$
while the decay $\jpsi\ar\GG$ decay is generated according to phase space.
\section{Event selection}
The data sample used for this analysis consists of $(14.00\pm 
0.56)\times 10^6~\psp$ events ~\cite{moxh} collected with the BESII
detector at the center-of-mass energy $\sqrt s=M_{\psp}=3.686$ GeV. 
Events with
two charged tracks and two photons are selected. 
Each charged track is required to be well fitted by a helix and to have a 
polar angle, $\theta$, within the fiducial region $|\cos\theta|<0.8$.
To ensure that tracks originate from the interaction region, we require
that $V_{xy}=\sqrt{V_x^2+V_y^2}<2$ cm and $|V_z|<20$ cm, where $V_x$, $V_y$,
and $V_z$ are the $x, y$, and $z$ coordinates of the point of closest
approach of each charged track to the beam axis. In addition, the
momentum for each charged track has to be less than 0.5 GeV/c. 

A neutral cluster is considered to be a photon candidate if it is
located within the BSC fiducial region ($|\cos\theta|<0.75$), the
energy deposited in the BSC is greater than 0.5 GeV, the first hit
appears in the first 6 radiation lengths, the angle between the
cluster and the nearest charged track is more than $15^\circ$, and
the angle between the direction of cluster development and the
direction of the photon emission is less than $40^\circ$.

A four constraint (4-C) kinematic fit under the $\psp\ar\pp\GG$
hypothesis is performed, and the $\chi^2$ of this fit is required to
be less than 9. Figure~\ref{mpipi} shows the distribution of mass
recoiling against the $\pp$ for the surviving events. The MC indicates the
detection efficiency for this channel is 16.0\%.
\bfg[htpb]
\centerline{\psfig{file=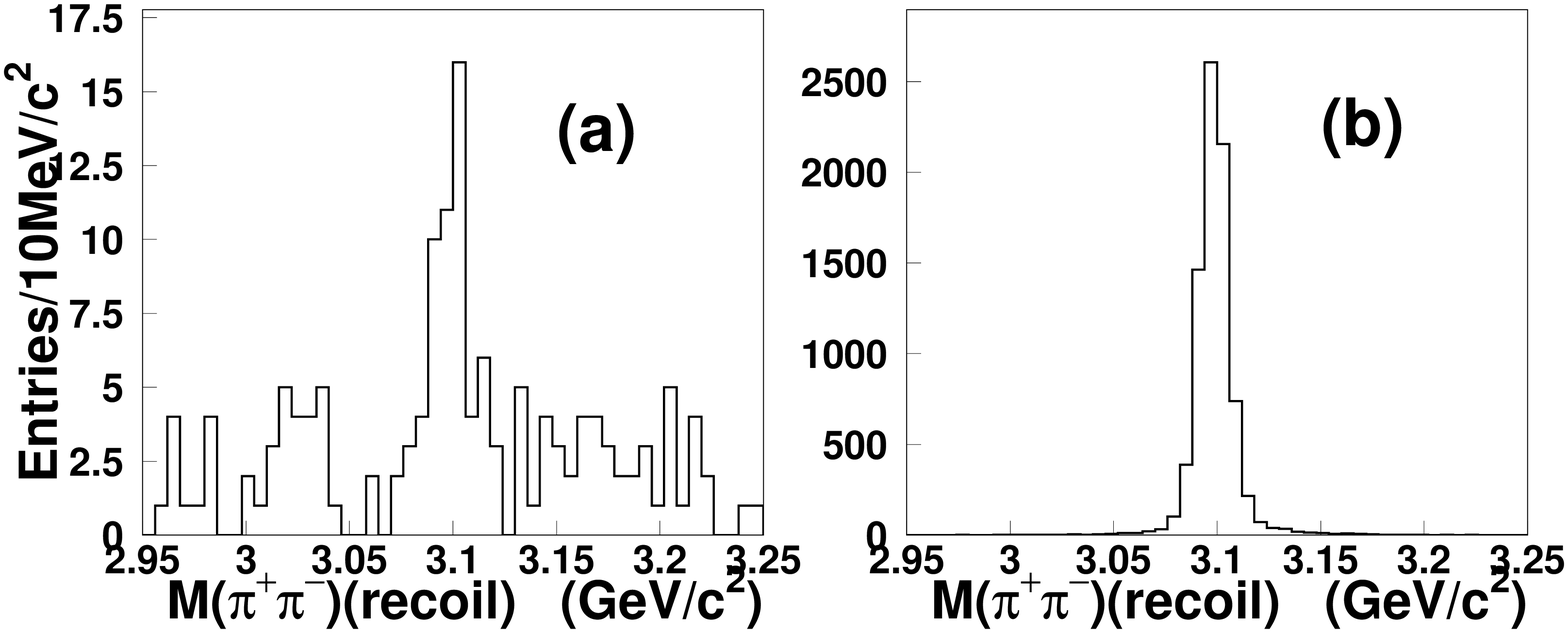,width=9cm,height=5.5cm}}
\caption{The distribution of mass recoiling against $\pp$ after event
selection. (a) data and (b) MC.}
\label{mpipi}
\efg
\section{Peaking background study}
The main peaking backgrounds come from $\psipto,\jpsi\ar neutral$, in
particular decays in which the final state contains three photons such
as $\jpsi\ar\gamma$ P (P denotes $\pi^0,\eta,\etap$), $P\ar\GG$.  In
addition, some background arises from five photon final states such as
$\jpsi\ar\gamma f_2(1270),f_2(1270)\ar\pi^0\pi^0$, etc.  The
efficiencies for other potential background channels in which the
number of photon is more than five, for example,
$\jpsi\ar\getap,\etap\ar\pi^0\pi^0\eta,\eta\ar\GG$, are found to be on
the order of $\sim$ 0.01\% and therefore negligible. 
Taking account the measured branching ratios of ${\cal
B}(\jpsi\ar\gpi,\geta,\getap)$~\cite{chenhx} and the combined
branching ratios of ${\cal B}(\jpsi\ar X,X\ar\pi^0\pi^0)$ (here X
denotes $f_2(1270),f_0(1500)$ and $f_0(1710)$)~\cite{gpipi}, the
normalized numbers of background events are determined and listed in 
Table~\ref{bkg}.  The contribution from other potential
background channels such as $\jpsi\ar\gamma f_2(1810),\gamma
f_0(2020),\gamma f_2(2150)$, and $\gamma f_4(2050)$ should not be
larger than that from $\jpsi\ar\gamma f_2(1270)$ according to the
partial wave analysis (PWA) result in
Ref.~\cite{gpipi}. Conservatively, $2.0\pm 1.0$ events are estimated
as the background from these four channels.  
\btbl[h]
\caption{Peaking background events from different channels. The ``Sum of four
  channnels''  includes
$\gamma f_2(1810), \gamma f_0(2020),\gamma f_2(2150)$, and $\gamma
f_4(2050)$.}
\bcl
\doublerulesep 2pt
\begin{tabular}{l|c}\hline\hline
Peaking background& Number of events\\
($\psipto,\jpsi\ar$)&   \\\hline
$\gpi$&$8.3^{+1.7}_{-1.2}$\\
$\geta$&$16.4\pm 1.4$\\
$\getap$&$1.5\pm 0.2$\\
$\gamma f_2(1270)\ar\gamma\pi^0\pi^0$&$1.7\pm 0.3$\\
$\gamma f_0(1500)\ar\gamma\pi^0\pi^0$&negligible\\
$\gamma f_0(1710)\ar\gamma\pi^0\pi^0$&$0.4\pm 0.3$\\
Sum of four channels& $2.0\pm 1.0$\\\hline
Total&$30.4\pm 2.4$\\\hline
\end{tabular}
\label{bkg}
\ecl
\etbl
\section{Mass spectrum fit}
A fit with a Breit-Wigner function convoluted with a Gaussian resolution 
function yields the number of $\jpsi$ events, $33.4\pm 6.6$. Here, the 
function to describe the background shape is a second-order polynomial.
Figure~\ref{fit} shows the fit results. 
\bfg[htpb]
\centerline{\psfig{file=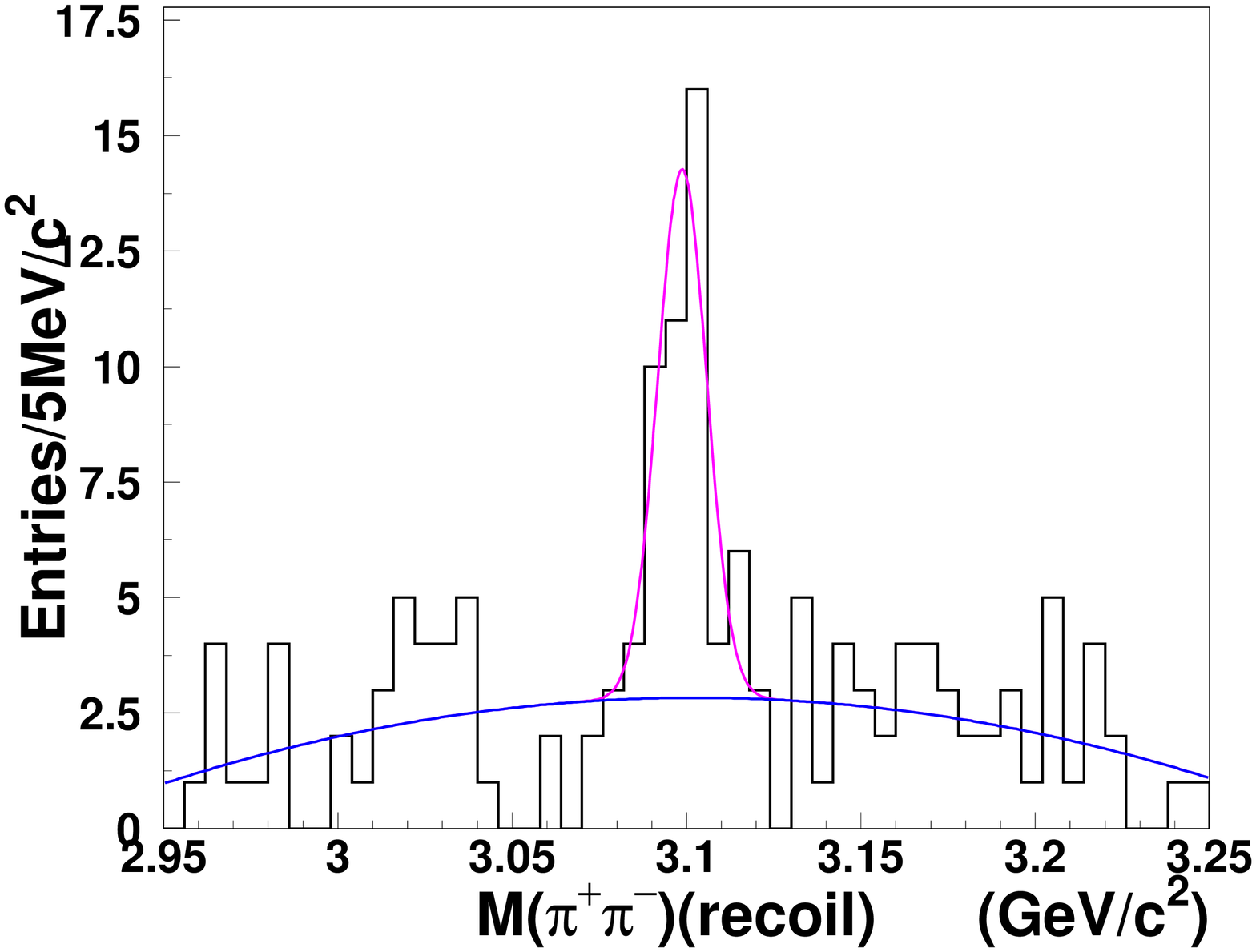,width=8cm,height=6cm}}
\caption{The distribution of mass recoiling from the $\pp$ after event
selection. The curves shows the fit described in the text.}
\label{fit}
\efg
\section{Systematic error}
The systematic error on this branching ratio measurement includes the
uncertainties in the MDC tracking efficiency, photon efficiency, kinematic 
fit, binning, fit range, background shape, the number of $\psi(2S)$ events, 
etc.
\subsection{MDC tracking efficiency and photon efficiency}
For charged tracks, the uncertainty of the tracking efficiency is determined 
by comparing data and MC, and an error of 2\% is found for each
track~\cite{simbes}. A similar comparison has also been performed for
photons, and the difference is also about 2\% for a single
photon~\cite{simbes}.
\subsection{Kinematic fit}
The final state in the investigated channel is the same as in
$\psp\ar\ppp$ where
an error of 6\% is given due to the uncertainty from kinematic fit~\cite{wangz}
Similarly, the same error is assumed in this analysis due to the uncertainty 
from the 4-C fit. 
\subsection{Binning, fit range and background shape}
When the bin width is changed from 25 MeV/$c^2$ to 50 MeV/$c^2$ and
the fit range is changed from 3.2 - 3.7 GeV/$c^2$ to 3.1 - 3.7
GeV/$c^2$, the differences of the numbers of the fitted signal events
is 5.2\%. This is considered as a systematic error.  The uncertainty
caused by using a different background shape is negligible.
\subsection{Total systematic error}
The total systematic error, determined by the sum of all sources
added in quadrature, is listed in Table~\ref{tot}. 
\btbl[h]
\caption{Systematic error(\%).}
\bcl
\doublerulesep 2pt
\begin{tabular}{l|c}\hline\hline
Source&(\%)\\\hline
Track efficiency&4\\
Photon efficiency&4\\
Kinematic fit&6\\
Binning \& fit range&5.2\\
Background shape& negligible\\
${\cal B}(\psipto)$& 1.9\\
$\psp$ total number&4\\\hline
Total &10.7\\\hline
\end{tabular}
\label{tot}
\ecl
\etbl
\section{\boldmath Upper limit of ${\cal B}(\jpsi\ar\GG)$}
The signal region for $\jpsi\ar\GG$ is taken from 3.08 to 3.12
GeV/$c^2$ in $\pp$ recoiling mass distribution, which is $\jpsi$ mass
plus and minus three standard deviation in the mass resolution. The
total number of events in the signal region is 52. The peaking background is 
30.4 and the smooth background is 52-33.4=18.6 according to fit result. 
Therefore, the expected number of total background
events is 49. With the Bayesian method~\cite{bayes}, the upper limit
on the number of $\jpsi\ar\GG$ events is estimated to be 15.74 at the
90\% confidence level, in which the systematic errors of the signal
detection efficiency (10.7\%) and the background expectation (5.0\%,
this uncertainty is obtained by 2.4 events from peaking background 
out of the total 49 events)
have been taken into account with the assumption that these two errors
are independent of each other. Therefore, the upper limit on ${\cal
B}(\jpsi\ar\GG)$ is \beqr {\cal
B}(\jpsi\ar\GG)&<&\frac{N_{obs}^{UL}}{N_{\psp}\cdot{\cal B}
(\psipto)\cdot\epsilon}\nonumber\\ &=&\frac{15.74}{14M\cdot
31.8\%\cdot 16.0\%}\nonumber\\ &=&2.2\times 10^{-5}\nonumber,\\ \eeqr
where $N_{obs}^{UL}$ is the upper limit on the number of events for
$\jpsi\ar\GG$, $N_{\psp}$ is the total number of $\psp$ events, and
$\epsilon$ is the detection efficiency.
\section{Summary}
The upper limit for the branching ratio ${\cal B}(\jpsi\ar\GG)$ at the
90\% confidence level has been 
measured via $\psipto$. Our upper limit for the C-violating decay is about 20 
times lower than previous measurements. It indicates that there is no obvious 
C-parity violation.
\section{Acknowledgments}
The BES collaboration thanks the staff of BEPC and computing center
for their hard efforts. This work is supported in part by the National 
Natural Science Foundation of China under contracts Nos. 10491300, 10225524, 
10225525, 10425523, the Chinese Academy of Sciences under contract No. KJ
95T-03, the 100 Talents Program of CAS under Contract Nos. U-11, U-24, U-25, 
and the Knowledge Innovation Project of CAS under Contract Nos. U-602, U-34
(IHEP), the National Natural Science Foundation of China under Contract No.
10225522 (Tsinghua University), the Swedish research Council (VR), and
the Department of Energy under Contract No.DE-FG02-04ER41291 (U Hawaii).
      
\end{document}